# ASSOCIATION RULE MINING BASED ON TRADE LIST

Ms. Sanober Shaikh[1] Ms. Madhuri Rao[2]

[1]Department of Information Technology, TSEC, Bandra (w), Mumbai
s.sanober1@gmail.com
[2]Department of Information Technology, TSEC, Bandra (w), Mumbai
my_rao@yahoo.com

## ABSTRACT

*In this paper a new mining algorithm is defined based on frequent item set. Apriori Algorithm scans the database every time when it finds the frequent item set so it is very time consuming and at each step it generates candidate item set. So for large databases it takes lots of space to store candidate item set .In undirected item set graph, it is improvement on apriori but it takes time and space for tree generation. The defined algorithm scans the database at the start only once and then from that scanned data base it generates the Trade List. It contains the information of whole database. By considering minimum support it finds the frequent item set and by considering the minimum confidence it generates the association rule. If database and minimum support is changed, the new algorithm finds the new frequent items by scanning Trade List. That is why it's executing efficiency is improved distinctly compared to traditional algorithm.*

## 1. INTRODUCTION

Mining Association rule is very important field of research in data mining. The problem of mining Association rule is put forward by R.S Agarwal first in 1993. Now the Association rules are widely applied in E-commerce, bank credit, shopping cart analysis, market analysis, fraud detection, and customer retention, to production control and science exploration. etc. [1]

Now a days we will find many mining methods for finding the frequent item set such as Apriori algorithm, Frequent Pattern-Tree algorithm etc. Apriori algorithm's disadvantage is it generates lot of candidate itemsets and scans database every time. If database contains huge number of transactions then scanning the database for finding the frequent itemset will be too costly and it generates a lot of candidates. Next FP-Tree algorithm's advantage is it does not produce any candidate items but it scans database two times in the memory allowed. But when the memory does not meet the need, this algorithm becomes more complex. It scans the database more than two times and the I/O expenses will increase [2]. That is why there is need to design an efficient algorithm which updates, protects and manages the association rule in large transactional database. So far many researchers made analysis and research for how to efficiently update the association rules and put forward corresponding algorithm. There are two instances in the problem of Association Rule update. The first instance is when the database is changed then how to find frequent item sets. FUFIA Algorithm is the representational updating method for this problem. The second instance is when the minimum support is changed then how to find frequent items sets. IUA algorithm is the representational updating method for this problem. These updating algorithms have both advantages and disadvantages. This paper proposes a dynamic algorithm of frequent mining based on undirected item set graph which scans the database only once and then saves the information of original database in undirected item set graph and finds the frequent item sets directly from the graph. It does not generate any candidate items. When database and minimum support is changed, the algorithm rescans the undirected item set graph to get the new frequent item sets.[3]

## 2. BASIC CONCEPT OF ASSOCIATION RULE

Association rule finds interesting associations and/or correlation relationships among large set of data items. Association rule shows attribute value conditions that occur frequently together in a given dataset. A typical and widely-used example of association rule mining is Market Basket Analysis.

For example, data are collected using bar-code scanners in supermarket. Such 'market basket' databases consist of a large number of transaction records. Each record lists all items bought by a customer on a single purchase transaction. Managers would be interested to know if certain groups of items are consistently purchased together. They could use this data for adjusting store layouts (placing items optimally with respect to each other), for cross-selling, for promotions, for catalog design and to identify customer segments based on buying patterns.

Association rules do not represent any sort of causality or correlation between the two item sets The problem of mining association rules can be described as below: if $I = \{I_1, I_2, \ldots I_n\}$ is the set of items. Suppose D is database transaction set and each transaction T contains set of items, such that $T \subseteq I$. Each transaction has identifier called as TID i.e. transaction id. Suppose A is a set of items and transaction T is said to contain A only if $A \subseteq T$.

Association rule is an implication like as $A \Rightarrow B$ in which A, B $\subset$ I and $A \cap B = \emptyset$. [6]

Definition of support: The support is the percentage of transactions that demonstrate the rule. An item set is called frequent if its support is equal or greater than an agreed upon minimal value – the support threshold. [8]

Definition of Confidence: Every association rule has a support and a confidence.

An association rule is of the form: $X \Rightarrow Y$.

$X \Rightarrow Y$: if someone buys X, he also buys Y.

The confidence is the conditional probability that, given X present in a transition, Y will also be present. Confidence measure, by definition:

Confidence($X \Rightarrow Y$) = support(X, Y)/ support(X)

The aim of association rule is to find all association problems having support and confidence not less than given threshold value. For the given support i.e. minsupp, if the item set of D's support is not less than minsupp, then it can say that D is the frequent item set.

## 3. FINDING FREQUENT ITEM SETS

First step is to scan the database. It makes each item as a node and at the same time it makes the supporting trade list for each node. Supporting trade list is a binary group T= {Tid, Itemset} (where Tid is transaction id and Itemset is trade item set). Given database that includes five items and nine transactions (shown in table one). Suppose that minimum support minsupp is two. Table two contains the information of support trade list of table one.

With this Trade List directly we will get information of which items are appearing in which transactions. So here number of transactions related to that item will decide count of that item. So we have count of I1 as 6 as shown in Table 2. Similarly we will get the count of all the items in the database. Now after considering the minimum support from user we will compare that minimum support with the count. If it is greater those will be considered as frequent-1 item set.

Table 1: A Store Business Data

| TID | The List Of Item ID |
|---|---|
| T100 | I1,I2,I5 |
| T200 | I2,I4 |
| T300 | I2,I3 |
| T400 | I1,I2,I4 |
| T500 | I1,I3 |
| T600 | I2,I3 |
| T700 | I1,I3 |
| T800 | I1,I2,I3,I5 |
| T900 | I1,I2,I3 |

Table 2: Trade List of Commodity Item

| Commodity Item | Support Trade List |
|---|---|
| I1 | T100,T400,T500,T700, T800, T900 |
| I2 | T100,T200,T300,T400, T600,T800, T900 |
| I3 | T300,T500,T600,T700, T800,T900 |
| I4 | T200,T400 |
| I5 | T100,T800 |

In next step for finding frequent itemset do intersection of I1 and I2. In result if we will get some transactions we will get common then it means that the item is related to other transaction also. Count the numbers of those common transactions that will give the count of those two items that are bought together that many numbers of times. Example I1$\cap$ I2 will get the count as 4 that means I1 and I2 are together 4 number of times in the database. Compare this with minimum support. Then we will get frequent-2 itemset. Similarly the procedure is iteratively applied.

### 3.1.3. Optimize Mining Algorithm Based On Undirected Item Sets Graph

The algorithm in this paper uses the search strategy of Depth first- Search to set universal undirected item graph. The specific steps are shown as follows:

Select a node $V_i$ from node set V. If the number of times $V_i$ appears in the database is not less than the minimum support minsupp, then {Vi} will belong to the item in frequent 1-item set. If count of node $V_i$ adjacent to node $V_j$'s side is not less than support S, then { $V_i$, $V_j$ } will belong to the item in frequent 2-iterm set. When there are three nodes in undirected item set graph and count of each side of the node is not less than minimum support minsupp, these three nodes < $V_k$,$V_m$,$V_n$> will belong to frequent 3-item set. When there more than three nodes in undirected item sets graph then count of each side of the node should not be less than minimum support minsupp and all the subset of these n nodes should be frequent.

## 3.2. Updating Trade List

When database and minimum support i.e. minsupp is changed the Trade List should be changed accordingly. If we want to add some new items to the database, then Trade List is updated accordingly.

### 3.2.1. Database Affair Changed

For example, when a new item T910 is added to table one; the result is as shown as in table three.

Table 3: The New Data in a Store

| TID | The list of items |
|---|---|
| T100 | I1,I2,I5 |
| T200 | I2,I4 |
| T300 | I2,I3 |
| T400 | I1,I2,I4 |
| T500 | I1,I3 |
| T600 | I2,I3 |
| T700 | I1,I3 |
| T800 | I1,I2,I3,I5 |
| T900 | I1,I2,I3 |
| T910 | I1,I4 |

A new item T910 have added at this time. So the arisen number of side <$I_1$, $I_4$> is two. As shown in fig.1, frequent 1-item set is L1= {I1, I2, I3, I4, I5};
frequent 2-item set is L2={{ I1 , I2},{ I1 , I3 },{ I1 , I5},{ I2 , I3 },{ I2 , I4 }, {I2, I5}, {I1, I4 }}; frequent 3-item set is L3={{ I1, I2, I3 },{ I1, I2, I5},{ I1, I2, I4}}.

### 3.2.2 Minimum support changed

For example, when the minimum support minsupp is three, frequent 1-item set={I1, I2 , I3 }; frequent 2- item is L2={{ I1, I2},{ I1 , I3},{ I2 , I3 }}.

# 4. RESULTS
## 4.1 Results of Apriori Algorithm

Fig1: Frequent Item Set with Apriori Algorithm with database shown in Table 1

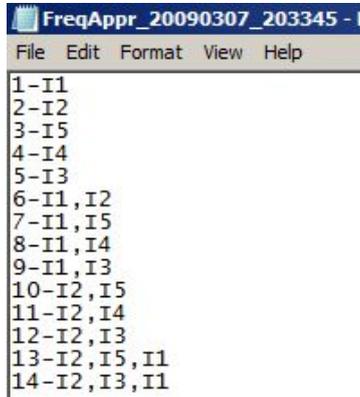

## 4.2 Results of Trade

Fig 2: Main Form

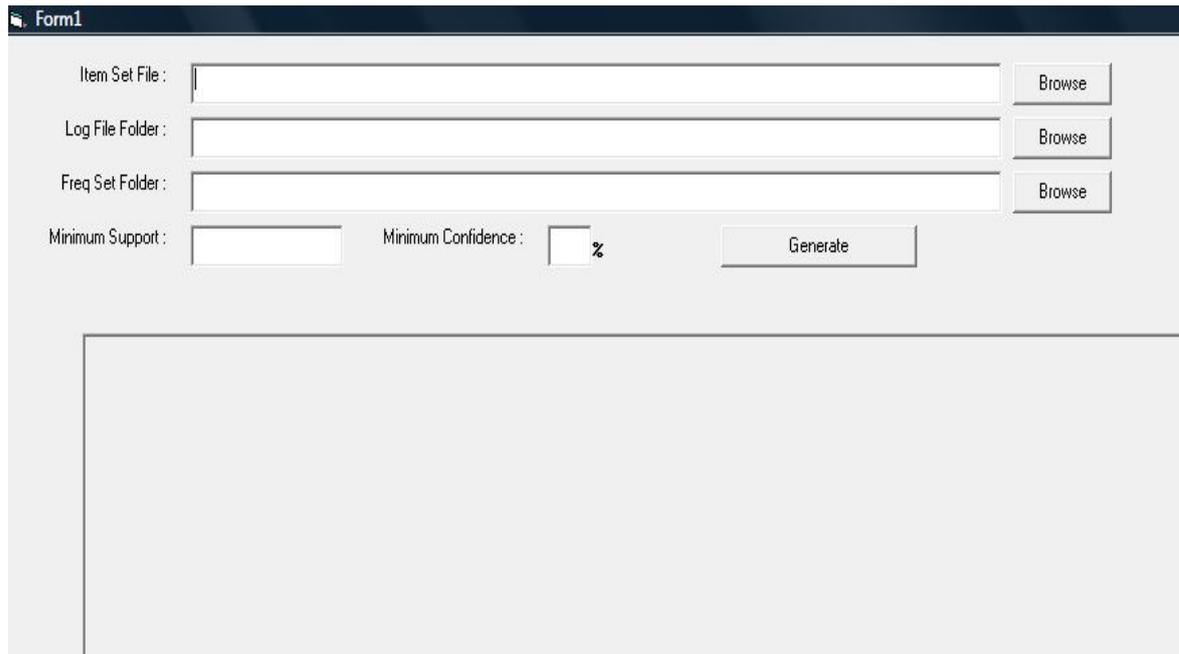

In Fig 2form the first i.e. Item Set File asks for the database from which you want to retrieve the frequent items. Here for input of Item set file one .isf file is made as shown in Fig 3. In that file the code for connectivity with database is made. Through the code the database is converted to a text file. In the first line write name of .isf file that will be converted to a format which the code will accept.

Fig 3: Item Set File

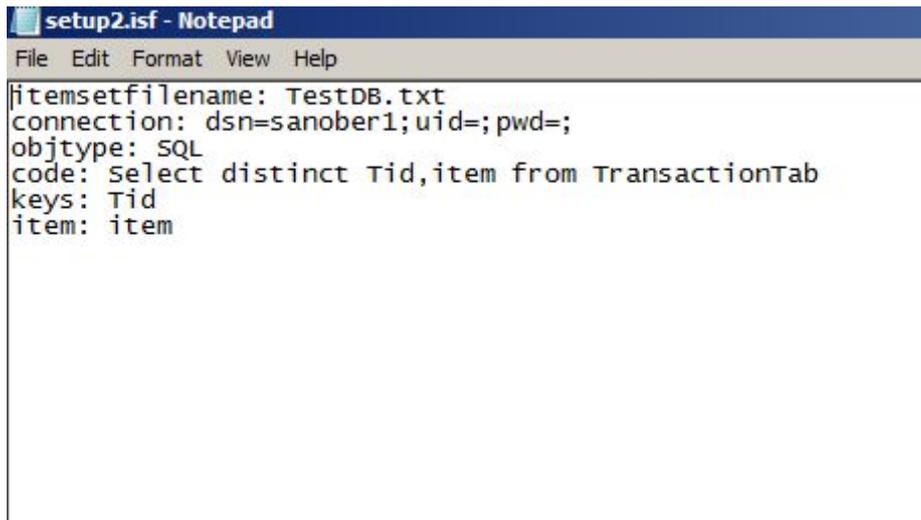

When we will click on Generate button in Fig 2, Trade list is made from which we can come to know how many number of items are present in input database as shown in Fig 4.

<div align="center">Fig 4: Trade List</div>

```
TradeList_20100226_114147.log - Notepad
File  Edit  Format  View  Help
I1  = T1,T2,T8,T12,T14,T15,T16,T22,T33,T34
I4  = T1,T2,T6,T13,T17,T19,T21,T22,T24
I5  = T1,T2,T4,T9,T11,T22,T23,T31,T35
I6  = T1,T2,T3,T5,T6,T10,T18,T19
I7  = T1,T4,T12,T14,T22
I11 = T1,T3,T4,T9,T11,T12,T14,T22
I19 = T1,T12,T13,T14
I23 = T1,T14
I25 = T1,T4,T11
I2  = T2,T3,T6,T8,T10,T13,T14,T15,T17,T19,T22,T24,T31
I3  = T2,T3,T9,T10,T12,T15,T20
I8  = T3,T6,T9,T12,T13,T19,T21,T31
I10 = T3,T6,T7,T8,T9,T10,T15,T17,T23,T25,T37
I13 = T3,T10,T11,T12,T14
I14 = T3,T5,T6,T8,T17,T19,T26,T30
I20 = T3,T7,T8,T11,T21,T25
I30 = T3,T25
I9  = T4,T12,T13,T14,T20,T24,T29,T36
I15 = T4,T7,T8,T10,T20,T28,T31
I9  = T4,T10,T11
I22 = T4,T5,T6,T8,T10,T19
I28 = T4,T5
I12 = T5,T6,T8,T10,T13,T18,T19,T20,T21,T26,T30,T33,T38
I18 = T5,T6,T8,T10,T18,T19,T20,T27,T29
I24 = T5,T6,T8,T18,T18,T19,T20,T27,T30
I30 = T5,T14,T18,T20,T27,T28,T30
I17 = T7,T12,T33
I26 = T8,T14
I20 = T10
I0  = T11,T13,T31
I6  = T11
I27 = T11,T20,T24,T29
I29 = T11
I21 = T14,T20,T24
I8  = T17
II6 = T20
I16 = T21,T26
I2  = T31
```

Then with the help of this Trade list we will get frequent items easily.

Here minimum support is 3. Now the count of each item is compared with minimum support. If count is greater than minimum support those items will be frequent item sets as shown in fig 5.

Fig 5: Frequent Items

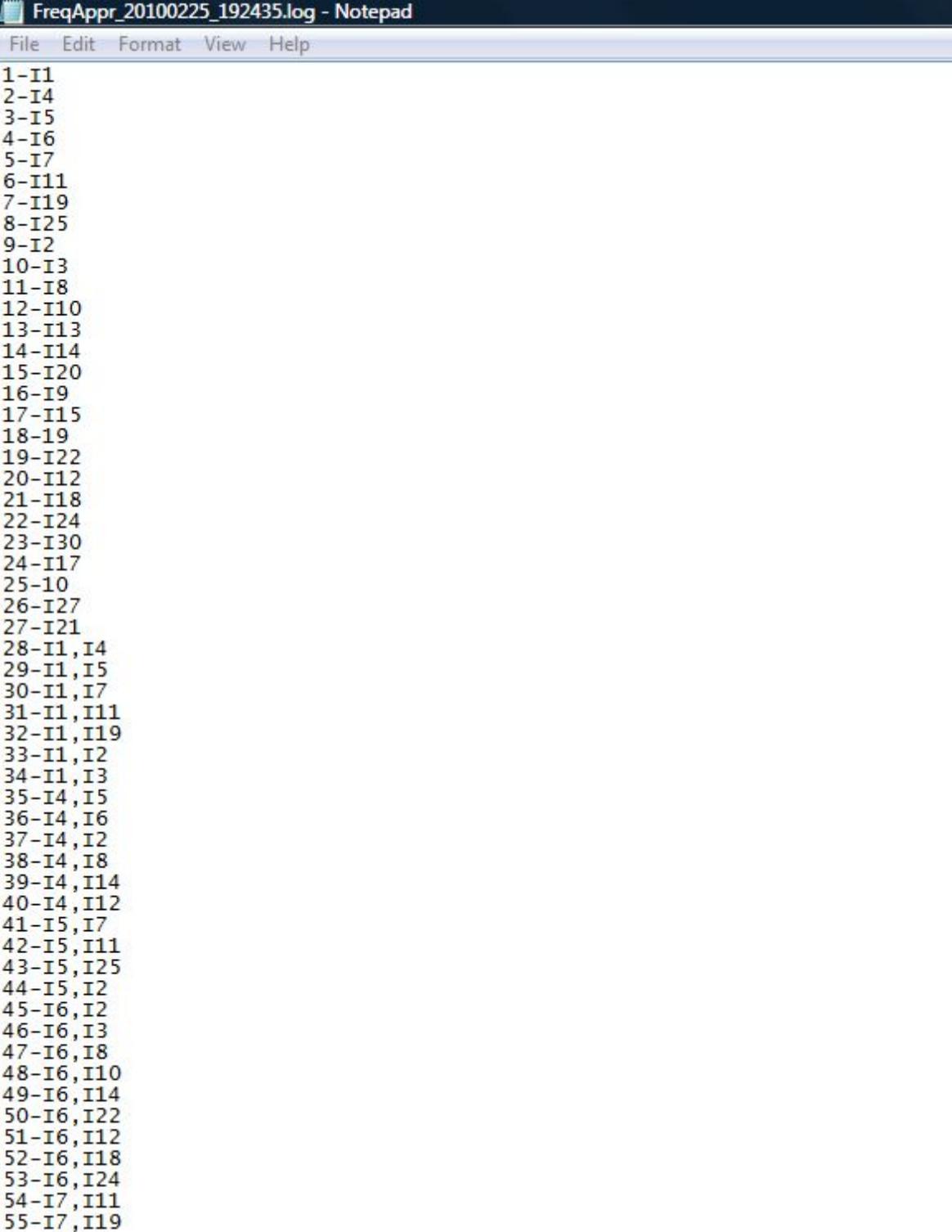

Fig 5 (cont): Frequent Items

```
FreqAppr_20100225_192435.log - Notepad
File  Edit  Format  View  Help
56-I7,I9
57-I11,I19
58-I11,I25
59-I11,I2
60-I11,I3
61-I11,I8
62-I11,I13
63-I11,I9
64-I19,I9
65-I2,I3
66-I2,I8
67-I2,I10
68-I2,I13
69-I2,I14
70-I2,I9
71-I2,I15
72-I2,I22
73-I2,I12
74-I2,I18
75-I2,I24
76-I3,I8
77-I3,I10
78-I3,I13
79-I8,I10
80-I8,I14
81-I8,I12
82-I10,I14
83-I10,I20
84-I10,I15
85-I10,I22
86-I10,I12
87-I10,I18
88-I14,I22
89-I14,I12
90-I14,I18
91-I14,I24
92-I9,I27
93-I9,I21
94-I15,I22
95-I15,I12
96-I15,I18
97-I22,I12
98-I22,I18
99-I22,I24
100-I12,I18
101-I12,I24
102-I12,I30
103-I18,I24
104-I18,I30
105-I24,I30
106-I4,I5,I1
107-I6,I2,I4
108-I2,I8,I4
109-I2,I14,I4
110-I2,I12,I4
```

Fig 5 (cont): Frequent Items

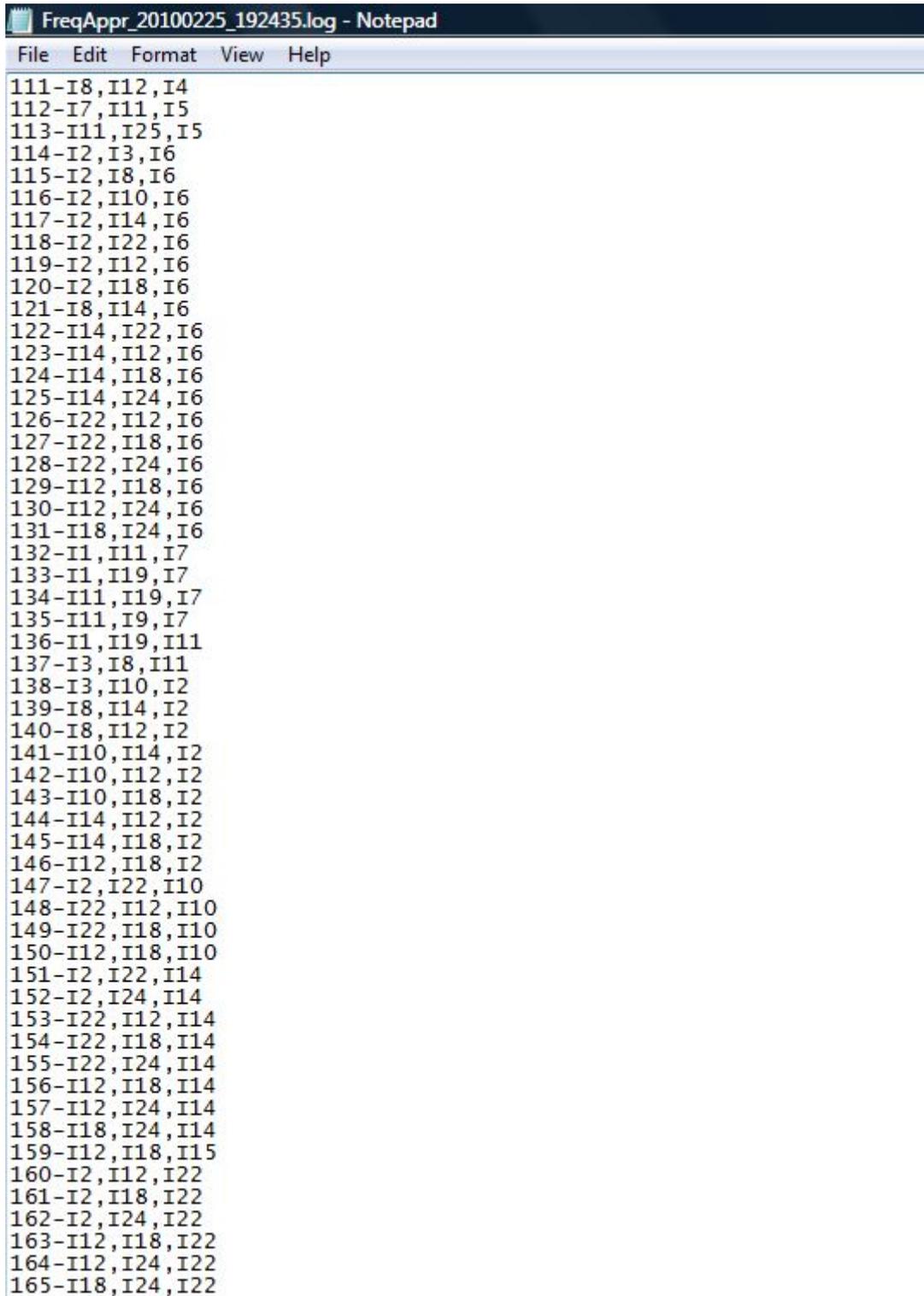

```
FreqAppr_20100225_192435.log - Notepad
File  Edit  Format  View  Help
166-I2,I24,I12
167-I18,I24,I12
168-I18,I30,I12
169-I24,I30,I12
170-I2,I24,I18
171-I24,I30,I18
172-I8,I12,I2,I4
173-I8,I14,I2,I6
174-I12,I18,I2,I6
175-I22,I12,I14,I6
176-I22,I18,I14,I6
177-I22,I24,I14,I6
178-I12,I18,I14,I6
179-I12,I24,I14,I6
180-I18,I24,I14,I6
181-I2,I12,I22,I6
182-I2,I18,I22,I6
183-I12,I18,I22,I6
184-I12,I24,I22,I6
185-I18,I24,I22,I6
186-I18,I24,I12,I6
187-I1,I19,I11,I7
188-I12,I18,I10,I2
189-I12,I18,I14,I2
190-I2,I12,I22,I10
191-I2,I18,I22,I10
192-I12,I18,I22,I10
193-I2,I12,I22,I14
194-I2,I18,I22,I14
195-I2,I24,I22,I14
196-I12,I18,I22,I14
197-I12,I24,I22,I14
198-I18,I24,I22,I14
199-I2,I24,I12,I14
200-I18,I24,I12,I14
201-I2,I24,I18,I14
202-I12,I18,I2,I22
203-I2,I24,I12,I22
204-I18,I24,I12,I22
205-I2,I24,I18,I22
206-I2,I24,I18,I12
207-I24,I30,I18,I12
208-I12,I18,I22,I14,I6
209-I12,I24,I22,I14,I6
210-I18,I24,I22,I14,I6
211-I18,I24,I12,I14,I6
212-I12,I18,I2,I22,I6
213-I18,I24,I12,I22,I6
214-I12,I18,I2,I22,I10
215-I12,I18,I2,I22,I14
216-I2,I24,I12,I22,I14
217-I18,I24,I12,I22,I14
218-I2,I24,I18,I22,I14
219-I2,I24,I18,I12,I14
220-I2,I24,I18,I12,I22
```

Confidence of each item is compared with minimum confidence given by user and strong association rule is formed. The items having confidence greater than or equal to minimum confidence, are stored in file shown in Fig 6.

Fig 6: Association Rule

```
I7->I1 = 80%
I19->I1 = 75%
I4,I5->I1 = 100%
I4->I2 = 77.78%
I6,I2->I4 = 60%
I2,I8->I4 = 60%
I2,I14->I4 = 60%
I2,I12->I4 = 60%
I8,I12->I4 = 100%
I8,I12,I2->I4 = 100%
I7->I5 = 60%
I11->I5 = 62.5%
I25->I5 = 100%
I7,I11->I5 = 60%
I11,I25->I5 = 100%
I6->I2 = 62.5%
I22->I6 = 66.67%
I6->I12 = 62.5%
I6->I18 = 62.5%
I2,I3->I6 = 75%
I2,I8->I6 = 60%
I2,I14->I6 = 60%
I2,I22->I6 = 75%
I2,I12->I6 = 60%
I2,I18->I6 = 75%
I8,I14->I6 = 100%
I14,I22->I6 = 75%
I14,I18->I6 = 75%
I14,I24->I6 = 60%
I22,I12->I6 = 80%
I22,I18->I6 = 80%
I22,I24->I6 = 75%
I6->I12,I18 = 62.5%
I12,I18->I6 = 71.43%
I8,I14,I2->I6 = 100%
I12,I18,I2->I6 = 75%
I22,I12,I14->I6 = 75%
I22,I18,I14->I6 = 75%
I22,I24,I14->I6 = 75%
I12,I18,I14->I6 = 75%
I12,I24,I14->I6 = 60%
I18,I24,I14->I6 = 75%
I2,I12,I22->I6 = 75%
I2,I18,I22->I6 = 75%
I12,I18,I22->I6 = 80%
I12,I24,I22->I6 = 75%
I18,I24,I22->I6 = 75%
I18,I24,I12->I6 = 66.67%
I12,I18,I22,I14->I6 = 75%
I12,I24,I22,I14->I6 = 75%
I18,I24,I22,I14->I6 = 75%
I18,I24,I12,I14->I6 = 75%
I12,I18,I2,I22->I6 = 75%
I18,I24,I12,I22->I6 = 75%
I18,I24,I12,I22,I14->I6 = 75%
```

Fig 7(cont): Association Rule

```
I7->I11 = 100%
I11->I7 = 62.5%
I7->I19 = 60%
I19->I7 = 75%
I7->I9 = 60%
I7->I1,I11 = 80%
I1,I11->I7 = 100%
I7->I1,I19 = 60%
I1,I19->I7 = 100%
I7->I11,I19 = 60%
I11,I19->I7 = 100%
I7->I11,I9 = 60%
I11,I9->I7 = 100%
I7->I1,I19,I11 = 60%
I1,I19,I11->I7 = 100%
I19->I11 = 75%
I25->I11 = 100%
I13->I11 = 80%
I1,I19->I11 = 100%
I3,I8->I11 = 100%
I19->I9 = 75%
I8->I2 = 62.5%
I13->I2 = 60%
I14->I2 = 62.5%
I22->I2 = 66.67%
I3,I10->I2 = 75%
I8,I14->I2 = 100%
I8,I12->I2 = 75%
I10,I14->I2 = 100%
I10,I12->I2 = 100%
I10,I18->I2 = 100%
I14,I18->I2 = 75%
I12,I18,I10->I2 = 100%
I12,I18,I14->I2 = 75%
I13->I3 = 60%
I20->I10 = 66.67%
I2,I22->I10 = 75%
I22,I12->I10 = 60%
I22,I18->I10 = 60%
I2,I12,I22->I10 = 75%
I2,I18,I22->I10 = 75%
I12,I18,I22->I10 = 60%
I12,I18,I2,I22->I10 = 75%
I22->I14 = 66.67%
I14->I12 = 75%
I14->I24 = 62.5%
I2,I22->I14 = 75%
I2,I24->I14 = 100%
I22,I12->I14 = 80%
I22,I18->I14 = 80%
I22,I24->I14 = 100%
I14->I12,I24 = 62.5%
I12,I24->I14 = 71.43%
I2,I12,I22->I14 = 75%
I2,I18,I22->I14 = 75%
```

Fig 7(cont): Association Rule

```
I2,I24,I22->I14 = 100%
I12,I18,I22->I14 = 80%
I12,I24,I22->I14 = 100%
I18,I24,I22->I14 = 100%
I2,I24,I12->I14 = 100%
I18,I24,I12->I14 = 66.67%
I2,I24,I18->I14 = 100%
I12,I18,I2,I22->I14 = 75%
I2,I24,I12,I22->I14 = 100%
I18,I24,I12,I22->I14 = 100%
I2,I24,I18,I22->I14 = 100%
I2,I24,I18,I12->I14 = 100%
I2,I24,I18,I12,I22->I14 = 100%
I27->I9 = 75%
I21->I9 = 100%
I22->I12 = 83.33%
I22->I18 = 83.33%
I22->I24 = 66.67%
I22->I2,I12 = 66.67%
I2,I12->I22 = 80%
I22->I2,I18 = 66.67%
I2,I18->I22 = 100%
I2,I24->I22 = 100%
I22->I12,I18 = 83.33%
I12,I18->I22 = 71.43%
I22->I12,I24 = 66.67%
I22->I18,I24 = 66.67%
I22->I12,I18,I2 = 66.67%
I12,I18,I2->I22 = 100%
I2,I24,I12->I22 = 100%
I22->I18,I24,I12 = 66.67%
I18,I24,I12->I22 = 66.67%
I2,I24,I18->I22 = 100%
I2,I24,I18,I12->I22 = 100%
I18->I12 = 77.78%
I24->I12 = 77.78%
I2,I24->I12 = 100%
I18,I24->I12 = 85.71%
I18,I30->I12 = 75%
I24,I30->I12 = 80%
I2,I24,I18->I12 = 100%
I24,I30,I18->I12 = 75%
I18->I24 = 77.78%
I24->I18 = 77.78%
I2,I24->I18 = 100%
I24,I30->I18 = 80%
I30->I24 = 71.43%
I2,I22->I12,I18 = 100%
I2,I22,I6->I12,I18 = 100%
I2,I22,I10->I12,I18 = 100%
I2,I22,I14->I12,I18 = 100%
```

## 5. CONCLUSION

In this project candidate items are not generated. The information of items of original database is saved in undirected item set graph. Then the information of frequent item set is found by searching trade list. The apriori scans the database too many times and generating candidates in each step. If we have huge amount of data then scanning such data and storage of huge amount of candidates is very difficult. Algorithm based on "A new association rule mining based on undirected itemset graph" having the disadvantage of tree generation. It takes time for generting tree. Now Trade list technique as compare to apriori and undirected itemset graph takes less amount of time and give the proper results.

## 6. ACKNOWLEDGEMENT

1. Ms Madhuri Rao(Guide)

2. Mr. Naushad Shaikh